\documentclass{aa}
\usepackage{graphicx}
\usepackage{txfonts}
\usepackage{hyperref}

\makeatother

\begin{document}

\title{Not all nitrogen-rich field stars originate from globular clusters} 

\author{E. I. Leitinger
      \inst{1,2}
      \and
      A. Miglio
      \inst{1,3,4}  
      \and
      J. Montalbán
      \inst{5,1,4}
      \and
      D. Massari
      \inst{3}
      \and 
      A. Bragaglia
      \inst{3}
      \and
      W. E. van Rossem
      \inst{1}
      \and
      K. Brogaard
      \inst{6,1}
      \and
      A. Mazzi
      \inst{1}
      \and
      J. S. Thomsen
      \inst{1}
      \and
      E. Willett
      \inst{4}
      }

\institute{Dipartimento di Fisica e Astronomia, Universit{\'a} degli Studi di Bologna, Via Gobetti 93/2, 40129 Bologna, Italy
     \and
         INAF - Osservatorio Astrofisico di Arcetri, Largo E. Fermi 5, 50125 Firenze, Italy
         \and
         INAF - Osservatorio di Astrofisica e Scienza dello Spazio di Bologna, Via Gobetti 93/3, 40129 Bologna, Italy         
         \and
         School of Physics \& Astronomy, University of Birmingham, Edgbaston, Birmingham B15 2TT, UK
         \and
         INAF - Osservatorio Astrofisico di Catania, Via S. Sofia, 78, 95123 Catania, Italy
         \and
          Stellar Astrophysics Centre, Department of Physics \& Astronomy, Aarhus University, Ny Munkegade 120, 8000 Aarhus C, Denmark
         }

\date{Received xxxx; accepted xxxx}

\abstract
{Globular clusters (GCs) are important tracers of the early Galactic assembly process, with part of their stars showing distinct chemical abundance patterns. When such stars are found in the Galactic field rather than within GCs, they are assumed to have originated from clusters. We expand the search for such chemically enriched stars in the \textit{Kepler} field, targeting stars located in the halo, thin and thick disc, to show the potential in using asteroseismology to link the inferred masses and hence, ages, with chemical abundances and kinematics. Using data from APOGEE DR17, \textit{Gaia} DR3, and the \textit{Kepler} mission, we identify primordial stars as those with chemical signatures typical of field stars, and enriched stars as those exhibiting strong nitrogen enrichment, with corresponding carbon and oxygen depletion. We present our sample of 133 red giant branch and core-He-burning stars, 92 of which have measured masses and inferred age estimations from asteroseismology. Of the 20 enriched stars identified, 13 have precise asteroseismic ages, of which a maximum of 3 are old enough ($> 8$ Gyr) to plausibly originate from globular clusters. The inferred asteroseismic ages indicate that most enriched stars found in the field appear too young to have originated from GCs; however, these apparently young ages are likely the result of assuming single-star evolution, rather than accounting for binary interactions or mergers. This points to alternative enrichment and evolutionary scenarios, such as mass transfer or coalescence, rather than a globular-cluster origin for most field nitrogen-rich stars. }

\keywords{stars: abundances -
            asteroseismology -
            globular clusters: general
           }

\maketitle
\nolinenumbers

\section{Introduction}
\label{sec:introduction}
One of the most distinctive features of globular clusters (GCs) is the presence of multiple stellar populations (MP), characterised by unique chemical abundance spreads. Of particular interest are nitrogen-rich stars, which often also exhibit enhancement in aluminium and sodium, but depletions in carbon, oxygen and sometimes magnesium \citep{2012Gratton,2018Bastian,2019Gratton,2020Meszaros,2022Milone}, although enrichment in aluminium does not necessarily coincide with nitrogen enrichment \citep{2022Fernandez}. These abundance patterns are characteristic of enriched stars in GCs, sometimes referred to as `P2' or `second-generation' stars. However, in this work, we refer to these stars simply as `enriched'. In comparison, GCs also contain `primordial' stars which show light element abundances consistent with field stars of the same metallicity. Currently, the formation and evolution history of MPs in GCs is uncertain, but the correlations and anticorrelations of light element abundances displayed by MPs are a distinct feature of GCs which allow us to identify field stars which may have once belonged to a GC. Although primordial stars are chemically indistinguishable from field stars of the same metallicity, when enriched stars are found wandering alone in the Galactic field, rather than bound to GCs or part of their extra-tidal structures, most theories assume they originated from GCs. Either these stars are the remnants of dissolved clusters or were ejected from surviving ones via tidal stripping and evaporation, processes that can disperse stars into the Galactic halo, disc, or bulge. For example, one of the most massive GCs in the Milky Way (MW), Terzan 5, is proposed to have an extensive history of expelling stars into the Galactic bulge and bar \citep{2009Ferraro, 2014Massari}, as supported by the chemo-dynamical tracing of an enriched star which was proposed to have been stripped from Terzan 5 approximately $350$ Myr ago \citep{2024Souza}. The identification of enriched stars in large-scale spectroscopic surveys, such as APOGEE, has opened new avenues for investigating the early dynamical and chemical history of the MW \citep{2017Schiavon}. From the present-day total stellar mass of the Galactic halo, $2-3\%$ is attributed to existing and intact GCs \citep{2018Forbes}, while current estimations on the percentage of stars which originated from a GC and now populate the Galactic halo range from $\sim 4\%$ to $\sim 70\%$ \citep{2021Horta,2023Belokurov}. The methods, data sets and abundances used to identify these enriched stars vary between works, contributing to this wide range of estimations.

For example, \cite{2016Martell} identified enriched stars using APOGEE DR12, determining that $2\%$ of the halo giants in their sample appeared to be enriched stars once belonging to GCs, and extrapolating that $13\%$ of all halo stars originated from GCs. Similarly, using SDSS-IV DR14 low-resolution spectroscopy, \cite{2019Koch} determined $\sim 11 \pm 1 \%$ of metal-poor ($-1.8 \leq $ [Fe/H] $\leq -1$) halo giants originated from disrupted GCs, with enriched stars contributing to approximately $3.7 \%$ of the Galactic halo. \cite{2020Hanke} associated field stars to GCs by studying their seven-dimensional chemo-dynamical information, tentatively estimating that $11.8 \pm 0.2 \%$ of halo stars once belonged to MW GCs, and classifying the stars as $\sim 80\%$ primordial and $\sim 20\%$ enriched. \cite{2021Horta} then concluded that disrupted GC stars contributed to $27.5^{+15.4}_{-11.5}\%$ of the field population at Galactocentric distances of $1.5$ kpc, with $16.8^{+10.0}_{-7.0}\%$ attributed to enriched stars. At a distance of $10$ kpc, these percentages dropped to $4.2^{+1.5}_{-1.3}\%$ of total stars and $2.7^{+1.0}_{-0.8}\%$ enriched stars. Finally, by analysing stars with high [N/O] ratios belonging to the oldest MW stellar population, \textit{Aurora}, \cite{2023Belokurov} estimated that $50 - 70\%$ of in-situ stars with [Fe/H] $\leq -1.5$ originated from bound GCs at high redshift. For stars with [Fe/H] $= -1$, their estimate reduced to $4-5\%$, suggesting that star formation in GCs contributed significantly to the early star formation of the MW. Building on this foundation, we have extended the search for chemically enriched stars into the Galactic halo and disc regions which were also observed by the \textit{Kepler} mission, with asteroseismic measurements that produce accurate masses and consequent age estimates of the individual stars. 

\vspace{-1em}
\section{Data sample selection and analysis}
\label{sec:data}

\begin{figure*}
    \centering
    \includegraphics[width=\textwidth]{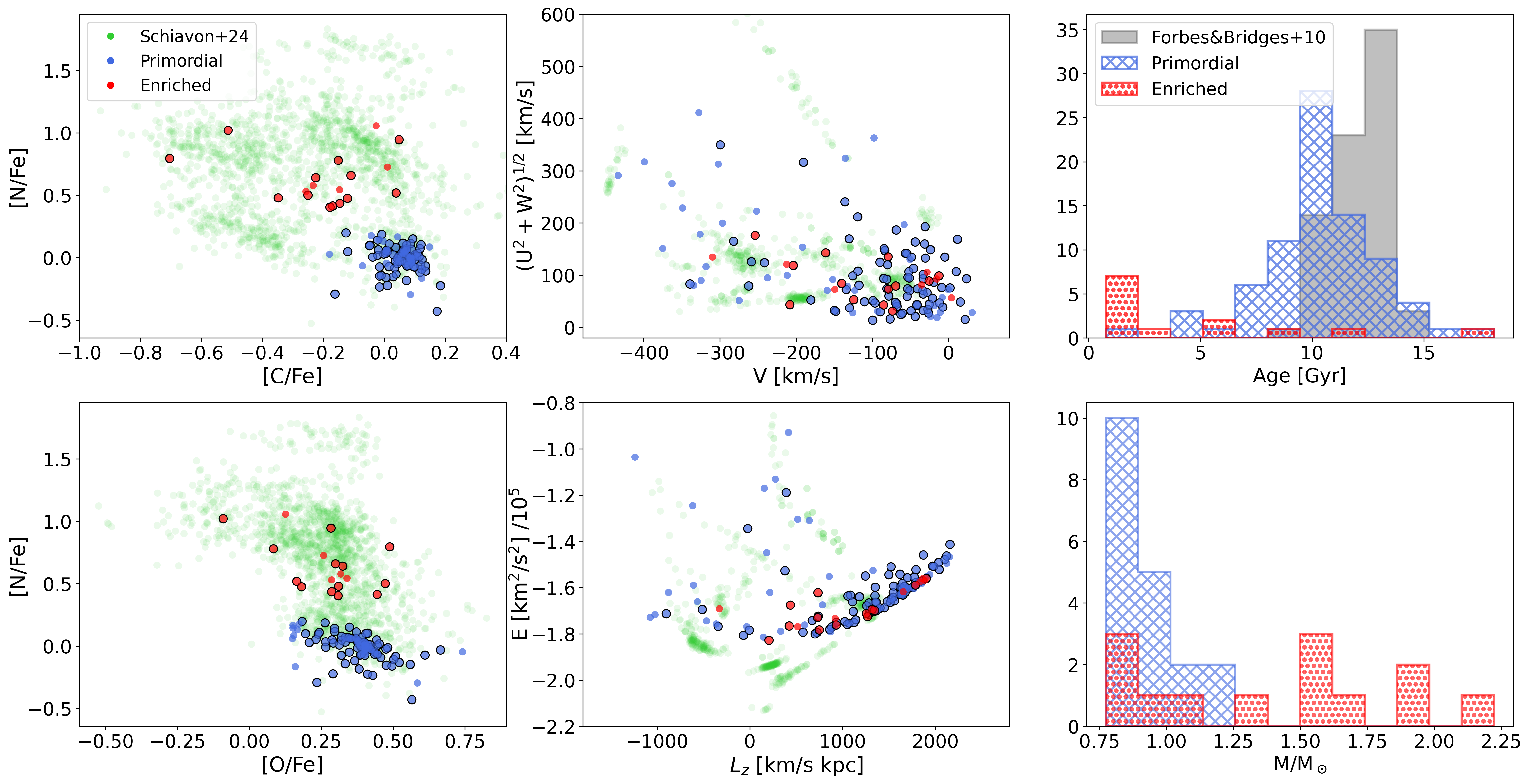}
    \caption{The primordial (blue) and enriched (red) field stars in the \textit{Kepler} sample, with the sample of GC stars identified by \cite{2024Schiavon} (green), both using APOGEE DR17 abundances. Stars in the \textit{Kepler} sample with ages determined through asteroseismology are shown with black borders. \textit{Upper left:} [N/Fe] vs. [C/Fe] anticorrelation, \textit{lower left:} [N/Fe] vs. [O/Fe] anticorrelation, \textit{upper middle:} Toomre diagram, \textit{lower middle:} integral of motion plot, \textit{upper right:} histogram of ages estimated from asteroseismology against GC ages from \cite{2010Forbes} (grey histogram), \textit{lower right:} histogram of masses derived from asteroseismology.}
    \label{fig:kepler_sample}
\end{figure*}

We searched for red giant stars in the \textit{Kepler} field with both chemical abundances from APOGEE DR17 and kinematic information from \textit{Gaia} DR3. The metallicity range of our selection is $-1.8 < \rm{[Fe/H]} < -0.5$, following the metallicity distribution observed in MW GCs, which peaks at around [Fe/H] $\sim -1.5$ \citep{2010Harris}. The models of \cite{2015Kruijssen} show that for two clusters of the same mass (i.e. $M \sim 10^5 M_\odot$), the cluster with [Fe/H] $\geq -0.7$ can lose 90\% of its initial mass, while the cluster with [Fe/H] $< -0.7$ can lose only 50\%, suggesting that the stars which originated from evaporated GCs, and are now lost in the halo, could be more metal-rich than the surviving GCs themselves. Our selection of red giant branch (RGB) and core-He-burning (CHeB) stars is based on the well-defined correlations and anti-correlations observed in MPs
\citep[e.g.][]{2009Carretta}. 

\subsection{APOGEE DR17}
\label{sec:data_APG}
We used data from the seventeenth data release of SDSS-IV \citep{2022Abdurrouf}, including stellar parameters and elemental abundances determined using the APOGEE Stellar Parameters and Chemical Abundance Pipeline (ASPCAP, \citealt{2016Garcia}). The \texttt{allStarLite} catalogue was used to supply the abundances, after removing stars from the sample based on the flags: \texttt{STAR\_BAD}, \texttt{TEFF\_BAD}, \texttt{LOGG\_BAD}, \texttt{VERY\_BRIGHT\_NEIGHBOR}, \texttt{LOW\_SNR}, \texttt{PERSIST\_JUMP\_NEG}, \texttt{PERSIST\_JUMP\_POS}, \texttt{SUSPECT\_RV\_COMBINATION} and duplicate observations removed according to the \texttt{EXTRATARG} bitmask. We do not remove stars with \texttt{PERSIST\_HIGH}, as persistence affects faint stars (\textit{H} $>$ 11) more than brighter stars \citep{2015Holtzman}, while our sample includes only red giant stars with \textit{H} $<$ 11. We also do not enforce an \texttt{ASPCAPFLAG} $== 0$ condition, as this restriction is not recommended by SDSS and solely removes primordial stars with weak CN features, mainly due to the \texttt{N\_M\_WARN} flag. Stars with abundance errors $> 0.25$ dex were removed for the elements C, N, O, Mg and Al, as well as for the metallicity values, for which we relied on the calibrated ASPCAP values. Red giants were then isolated using cuts in $\rm{T_{eff}} < 5300$ K, with SNR $>70$, following the same general criteria as \cite{2023Belokurov}, isolating the metal poor stars with [Fe/H] $< -0.5$. Additionally, we implement cuts in surface gravity to isolate the range of $1.5 <$ log(g) $< 3.5$ for the stars in our sample. The cut of log(g) $< 3.5$ follows the method of \cite{2023Belokurov} for selecting red giants, while a cut of $1.5 <$ log(g) represents the limit below which global asteroseismic observables ($\Delta \nu$, $\nu_{\mathrm{max}}$) become less reliable for inferring stellar properties such as mass and age, owing to increased uncertainties and less well-tested calibrations (see discussions in \citealt{2026Willett}, \citealt{2021Miglio} and \citealt{2021Montalban}).
\subsection{Gaia}
\label{sec:data_gaia}
We cross-matched stars common to \textit{Kepler} and APOGEE DR17 with \textit{Gaia} DR3 \citep{2016gaia1,2022gaia2} to obtain precise proper motions, radial velocities and parallaxes. We implemented cuts to ensure the sample contained reliable astrometry by removing stars with RUWE $>1.4$, \texttt{ipd\_frac\_multi\_peak} $>2$ and \texttt{ipd\_gof\_harmonic\_amplitude} $>$ exp $[0.18(G-33)]$, following the cuts specified by \cite{2021Fabricius} to remove non-single objects, as well as ensuring the value of \texttt{parallax\_over\_error} $>$ 5. We then applied the parallax zero-point correction to account for the known systematic offsets in \textit{Gaia} parallaxes, as characterised by \cite{2021Lindegren}, using the \texttt{zpt} package and only stars with reliable astrometric solutions (\texttt{astrometric\_params\_solved} $>$ 30) for the analysis. Previously, the work of \cite{2025Kane} identified candidate stars predicted to originate from GCs in the Galactic field, using a trained neural network on low-resolution \textit{Gaia} BP/RP spectra, from which they predicted [N/O] and [Al/Fe] abundances. We cross-matched this new abundance catalogue from \cite{2025Kane} with the \textit{Kepler} catalogue, adding another 29 stars to our sample with no abundances in APOGEE. We discuss this further in Appendix \ref{sec:appendixA}. 

Using astrometry and radial velocities from \textit{Gaia} DR3, we calculated the angular momentum ($L_z$) and energy ($E$) for each star using \texttt{galpy} and assuming a potential model from \cite{2017Mcmillan}. We then converted the 3D velocity components from the Galactic reference frame into the Local Standard of Rest (LSR) frame, using parallax, proper motions, and radial velocities from \textit{Gaia} DR3 to calculate the $U$, $V$, and $W$ components of velocity in the LSR frame. Using these parameters, we constructed the integrals of motion (IoM) and Toomre diagrams for all stars in our sample, as shown in the middle panels of Figure \ref{fig:kepler_sample}. We note that this model including non-axisymmetric terms may cause stars to clump on overdensities, however, we mainly reference the IoM plots in terms of whether specific stars have prograde or retrograde orbits, which are not affected by the overdensity clumping (see e.g. \cite{2025Woudenberg}, \cite{2025DeLeo}). In the upper middle panel of Figure \ref{fig:kepler_sample}, the Toomre diagram shows these stars are located within the halo, thin and thick disc. In the lower middle IoM plot, we see that most stars display orbital energies $E$ and $L_z$ values consistent with prograde orbits corresponding to in-situ formation, a small subset is also located at higher orbital energies with lower $L_z$, and some stars show retrograde orbits.
\subsection{\textit{Kepler} seismology}
\label{sec:data_seismo}
We considered red giant stars observed by the \textit{Kepler} mission \citep{2010Borucki} with  average seismic parameters  $\nu_{\mathrm{max}}$ and $\Delta \nu$  available in  \citealt{2018Yu}.  These parameters provide insight into the stellar structure, enabling precise measurements of stellar radii and masses, hence of stellar ages.
The asteroseismic parameters described above were used as inputs by \cite{2026Willett}, applying the Bayesian estimation code \texttt{PARAM} \citep{2006daSilva,2014Rodrigues,2017Rodrigues} using as constraints metallicity and $T_{\rm{eff}}$ from APOGEE (with minimum uncertainties of $0.05$ dex and $50$ K, respectively - see \citet{2026Willett} for details), $\nu_{\mathrm{max}}$ and either $\Delta \nu$ or luminosity in order to infer precise masses, distances, radii and consequent age estimations. This yielded  masses with associated uncertainties of $\sim 6\%$, while the age estimations have uncertainties of $\sim 23\%$. We note that age estimates for low-mass CHeB stars are affected by the poorly constrained efficiency of mass loss during the preceding RGB phase. In this work we follow \citet{2021Miglio} and \citet{2026Willett}, adopting a Reimers mass-loss parameter of $\eta = 0.2$. However, the true efficiency of mass loss likely depends on metallicity (\citealt{2024Brogaard}), so the derived ages for CHeB stars should be interpreted with caution. Additionally, for RGB stars, we sourced masses and ages from \cite{2021Montalban} who uses the Bayesian stellar parameter estimation tool \texttt{AIMS} with individual mode frequencies from asteroseismology, with uncertainties of $\sim 2\%$ for the masses and $\sim 11\%$ for the ages. We increased this sample by including newly analysed stars (Montalbán et al., (in prep.)), using the same method as \cite{2021Montalban} and \cite{2025Casali}. We preferentially use the masses and ages obtained through \texttt{AIMS}, if available. Of the 13 enriched stars in our sample, only KIC 8110538 shows a significant difference ($\sim 2.5 \sigma$ significance, with error propagation) in the calculated age between these methods, which is explored further in Appendix \ref{sec:appendixB}.
\subsection{Classification of primordial and enriched stars}
\label{sec:data_pops}
This work compares two main samples of stars in the \textit{Kepler} field: those which show chemical abundance patterns consistent with the enriched stars found in Galactic GCs, and a control sample which is representative of the primordial stars. Using the sample of cleaned and combined \textit{Kepler}, \textit{Gaia} DR3 and APOGEE DR17 data, we identified candidate GC stars, based on well known correlations and anticorrelations such as [C/Fe] vs. [N/Fe], [O/Fe] vs. [N/Fe] and [Mg/Fe] vs. [Al/Fe]. The abundance limits we have adopted to isolate primordial stars in the \textit{Kepler} sample are as follows: [N/Fe] $<$ 0.2, [Al/Fe] $<$ 0.1, -0.2 $<$ [C/Fe] $<$ 0.2, and [O/Fe] $>$ 0.1. We defined the candidates for enriched stars using a cut in nitrogen ([N/Fe]$>0.4$), as the separation between primordial and enriched stars occurs for each abundance distribution at around [N/Fe] = 0.4, which is most clearly seen for [O/Fe] vs. [N/Fe] in the lower left panel of Figure \ref{fig:kepler_sample}. This threshold aligns well with the distinction seen in MPs, where enriched stars display enrichment in nitrogen due to the products of hydrogen burning through the C-N-O cycle. To better constrain the selection, we incorporated additional cuts in abundances which are consistent with the signatures of enriched GC stars. These signatures include enrichment in aluminium ([Al/Fe] $\geq$ 0.1), with a depletion in carbon ([C/Fe] $\leq$ 0.2), due to the high-temperature proton-capture reactions which likely occurred in their progenitor environments. We further justify the specific cuts to separate stars into a primordial and enriched sample in Appendix \ref{sec:appendixPop}. The 113 primordial stars meeting the criteria are shown in blue for all panels of Figure \ref{fig:kepler_sample}, with the 20 enriched stars shown in red, alongside confirmed GC stars using APOGEE abundances, as identified by \cite{2024Schiavon} (green), to illustrate the general abundance trends observed in MW GCs. 

Although sodium abundances are available in APOGEE and there is a well known anticorrelation between [Na/Fe] and [O/Fe] for the MPs in GCs \citep{2009Carretta}, we did not use these Na abundances to classify the stars into populations, as the Na lines in the H band obtained from the infrared spectra of APOGEE cannot be measured reliably, which means this anticorrelation cannot be reproduced \citep{2019Masseron}.

\section{Discussion}
\label{sec:discussion}
The relative age distribution of GCs in the Galactic thick disc and halo was compiled by \cite{2010Forbes}, using ages obtained by multiple other works such as \cite{2009MarinFranch}, \cite{2008Dotter}, \cite{2005deAngeli}, \cite{2002Catelan} and \cite{1998Salaris}. The sample contained 76 Galactic GCs, in which the lowest age on this scale belongs to Eridanus ($8.9$ Gyr with metallicity [Fe/H] $= -1.2$), while the oldest is Lyngå 7 ($14.5$ Gyr with [Fe/H] = $-0.6$). In the upper right panel of Figure \ref{fig:kepler_sample} we show this distribution in grey, against the ages determined through asteroseismology for 92 stars in our sample, divided into primordial (blue) and enriched (red). The 79 primordial stars in the lower right panel of Figure \ref{fig:kepler_sample}, have masses typical of old field giants ($0.8 - 1.1 \rm{M_\odot}$). However, the resulting age distribution is then younger than GC stars and more consistent with the age distribution of thick disc field stars \citep{2020Helmi}. This sample also includes few apparently young stars ($<6$ Gyr), similar to previously flagged `overmassive' field stars with asteroseismic masses that are larger than expected for their metallicity and evolutionary stage (i.e. \cite{2021Montalban,2012Miglio}). These apparently young, overmassive stars have already been identified in open clusters \citep{2021Brogaard} and the [$\alpha$/Fe]-enhanced population \citep{2015Chiappini}, and are theorised to be a result of mass-transfer from a binary companion.

However, the most interesting result arises from the distribution of ages and masses for the 13 enriched stars with age estimations from asteroseismology, shown in red in the right panels of Figure \ref{fig:kepler_sample} and listed in Table \ref{tab:kepler_sample}. The masses derived from asteroseismology for the enriched stars span a much larger range than the primordial stars, with a median and standard deviation of $1.5 \pm 0.4 \rm{M_\odot}$. Consequently, the median age of the enriched sample is $1.9$ Gyr (between $0.8 <$ and $< 18.1$ Gyr), suggesting that the majority of enriched field stars in the \textit{Kepler} field are apparently too young to have originated from GCs (i.e. $< 8$ Gyr) and would therefore require an alternative enrichment origin. One such alternative is enrichment due to mass-transfer in a binary system (see, e.g. \cite{2019Trincado}, \cite{2025Kravtsov}, and references therein), which could reproduce the majority of the enrichment signatures seen in enriched GC stars, while also allowing for the primary stars/merger products to appear younger than those typically found in Galactic GCs. As the primordial and enriched samples follow similar distributions in the Toomre diagram and IoM, but not in mass and age, this could naturally be explained if the enriched sample contained mostly post-mass-transfer binaries. Most of the massive and therefore apparently young ($<8$ Gyr) enriched stars within our sample are classified as CHeB, as demonstrated in Table \ref{tab:kepler_sample}, with seven classified as CHeB stars and three as RGB stars. This provides additional evidence which supports a mass-transfer binary scenario, as the population synthesis simulations of \cite{2025Mazzi} demonstrated that post-mass transfer products are more likely to be CHeB stars than RGB stars in the \textit{Kepler} field. Within our enriched star sample, we already confirm that one star has been identified as a spectroscopic binary SB1 (KIC 10796857), as described further in Appendix \ref{sec:appendixB}. We aim to expand the sample of stars and thoroughly investigate potential signs of binarity in both the primordial and enriched samples. 

We must also consider the scenario in which the enriched stars of our sample formed within a young GC before being relocated to the field. To explore this case we will compare our enriched sample with the youngest known MW GCs. Within our sample are ten enriched stars with apparent ages $<8$ Gyr. Currently, $<2-3\%$ of Galactic GCs are shown to have ages $<8$ Gyr \citep{2009MarinFranch, 2013Vandenberg, 2023Massari}. This includes for example Terzan 7 ([Fe/H] averaged as $-0.45$) with an age of $7.75$ Gyr \citep{2013leaman}, the bulge/disc GC Palomar 1 ([Fe/H] $=-0.7$) with an age of $7.3$ Gyr \citep{2009MarinFranch}, the disc GC 2MASS-GC01 ([Fe/H] = $-0.85^{+0.07}_{-0.06}$) with an age of $7.22^{+0.93}_{-1.11}$ Gyr \citep{2025Massari}, and Whiting 1 ([Fe/H] $= -0.58 \pm 0.05$) with an age of $5.0 \pm 0.5$ Gyr \citep{2024Huang}. There has been no spectroscopic investigation of multiple stellar populations within any of these four young MW GCs, and therefore there are no enriched stars to compare with our sample. Then, there are the cases of the bulge fossil fragment candidate systems Terzan 5 and Liller 1, which are more complicated as they both contain multiple stellar populations, but with a bimodal distribution of metallicities accompanied by a spread in stellar ages. More specifically, the stars of both Terzan 5 and Liller 1 can be separated into an old, metal-poor population with metallicity peaks around [Fe/H] $\sim -0.5$ and a young, metal-rich population with [Fe/H] $\sim +0.3$ \citep{2023Crociati}. This is explained as being due to bursts of star formation activity, possibly through interactions with other bulge substructures, with the latest star formation events occurring around 4–5 Gyr ago in Terzan 5 and 1–2 Gyr ago for Liller 1. Focusing on the stars of Liller 1, the metal-poor group have apparent ages $> 10$ Gyr, while the metal-rich group have apparent ages between $1$ and $3$ Gyr \citep{2021Ferraro}. Therefore it is only the old, metal-poor stars in Liller 1 which are in a metallicity range compatible with our enriched sample. In terms of enriched stars found in Liller 1, \cite{2025Liptrott} classified primordial and enriched populations using APOGEE DR17 abundances, identifying one enriched star with [N/Fe] $= 0.82$ and [Fe/H] $= -0.59$, which is within the same metallicity and nitrogen abundance range as our enriched sample. However, this star is again in the metallicity regime of the older stars in Liller 1, not the younger stars. If we consider that our sample contains ten (out of 13) enriched stars with ages $<8$ Gyr, the likelihood that they originated from a young GC, and that this former GC still exists without having been dissolved, is very low. In the case of a young GC dissolving recently, there should be evidence of a cold stream, in which unbound stars are expected to follow a similar orbit \citep{2018Balbinot}. Furthermore, it would be expected that if very young GCs were dissolved in the MW, we would have discovered a large population of apparently young and massive primordial stars which also once belonged to the dissolved GC. We have indeed discovered young, massive primordial stars in our sample, but not to the same degree as the enriched stars, as shown in the right panels of Figure \ref{fig:kepler_sample}. The likelihood of an enriched star in our sample once belonging to a young GC is therefore very low for stars with asteroseismic ages between $6$ and $8$ Gyr, i.e., ages more consistent with the youngest GCs in the MW. This likelihood then depletes rapidly for lower ages, especially for the seven enriched stars with ages $< 2$ Gyr.

Now we will consider the case in which the asteroseismic ages derived for the enriched stars in our sample are not actually their true ages, but instead a result of mass-transfer in a binary system. That is, mass-transfer enriched the RGB or CHeB star of the binary system, causing it to become more massive and consequentially appear younger than it actually is. In this work we have proposed the idea that these binary systems exist in the field, but we can also consider the case in which they are formed within a GC before being ejected into the field. In the latter case, the mechanism ejecting the enriched stars from the GC to the field would need to be related to the binary itself, via dynamical encounters. If it was not related to the binary, we would have discovered more single-star counterparts in the enriched sample which have ages consistent with MW GCs. However, even when considering the age range of the youngest MW GCs mentioned above, only four stars in our sample have ages $> 6$ Gyr, while nine stars have ages $< 6$ Gyr. These numbers could potentially be the result of a small sample size, so a follow-up investigation on stars in the K2 and TESS fields is in preparation, using the same method. In addition, when we also consider that some enriched stars in our sample have masses greater than twice the turn-off, it becomes necessary for three stars to be involved in the evolution of the binary. The dynamical simulations of \cite{2020Fragione} show that while more than $15\%$ of field stars have at least two stellar companions, GCs only host tens of these systems, making the likelihood of this occurring much lower in GCs than in the field.

Our results have an impact on the previous estimations made on the fraction of GC stars assumed to have contributed to the field star population of the MW. A conservative fraction of only $23\%$ (3 out of 13) of the enriched stars in our sample exhibit ages consistent with GC formation, if we consider only the enriched stars with ages $>8$ Gyr, consistent with the age distribution of Galactic thick disc and halo stars \citep{2010Forbes}. Future investigations into nitrogen-rich field stars would benefit from both combining age estimations from asteroseismology to their samples, where available, as well as performing follow-up spectroscopic observations to obtain additional abundances and radial velocity measurements, to further identify and classify these seemingly young stars.

\section{Conclusions}
\label{sec:conclusions}

We have identified 133 metal-poor halo, thin and thick disc red giant stars in the \textit{Kepler} field, 92 of which have precise masses and estimated ages determined through asteroseismology. Using stellar abundances from APOGEE DR17, we chemically tag stars as either primordial or enriched, following the expected correlations and anticorrelations associated with Galactic GC stars. Of the 13 enriched stars with asteroseismic age estimations, we have identified only three stars consistent with the age range expected for MW GCs (older than $\sim 8$ Gyr) which could have undergone enrichment within a GC before being relocated into the field either via ejection or the dissolution of their GC. However, the majority of enriched field stars in our sample exhibit apparent ages that are too young for such a scenario to be plausible. This result presents a counter-claim to the assumption that all nitrogen-rich stars in the field must have originated from GCs, which has implications on the estimated fraction of GC stars that have contributed to the Galactic field, and it highlights the necessity of exploring alternative enrichment scenarios which could mimic the enrichments observed in GC stars, in particular binary interaction products.

\begin{acknowledgements}
We thank the referee for their useful comments which helped to improve this manuscript. EL thanks Elena Pancino for her discussions and support. EL, AM, and WEvR acknowledge support from the ERC Consolidator Grant funding scheme (project ASTEROCHRONOMETRY, \url{https://www.asterochronometry.eu}, G.A. n. 772293). Co-funded by the European Union (ERC-2022-AdG, {\em "StarDance: the non-canonical evolution of stars in clusters"}, Grant Agreement 101093572, PI: E. Pancino). Views and opinions expressed are however those of the author(s) only and do not necessarily reflect those of the European Union or the European Research Council. Neither the European Union nor the granting authority can be held responsible for them. A.Mazzi and JST acknowledge financial support from the ``MUR FARE Grant Duets CUP J33C21000410001''. Part of the calculations described in this paper were
performed using the University of Birmingham BlueBEAR HPC service (\url{http://www.birmingham.ac.uk/bear}). DM acknowledges financial support from PRIN-MIUR-22: CHRONOS: adjusting the clock(s) to unveil the CHRONO-chemo-dynamical Structure of the Galaxy” (PI: S. Cassisi)
\end{acknowledgements}

\bibliographystyle{aa} 
\bibliography{references}

\label{LastPage}

\begin{appendix}

\section{Population classification using the new abundance catalogue from \cite{2025Kane}}
\label{sec:appendixA}

\begin{figure}[h]
    \centering
    \includegraphics[width=\columnwidth]{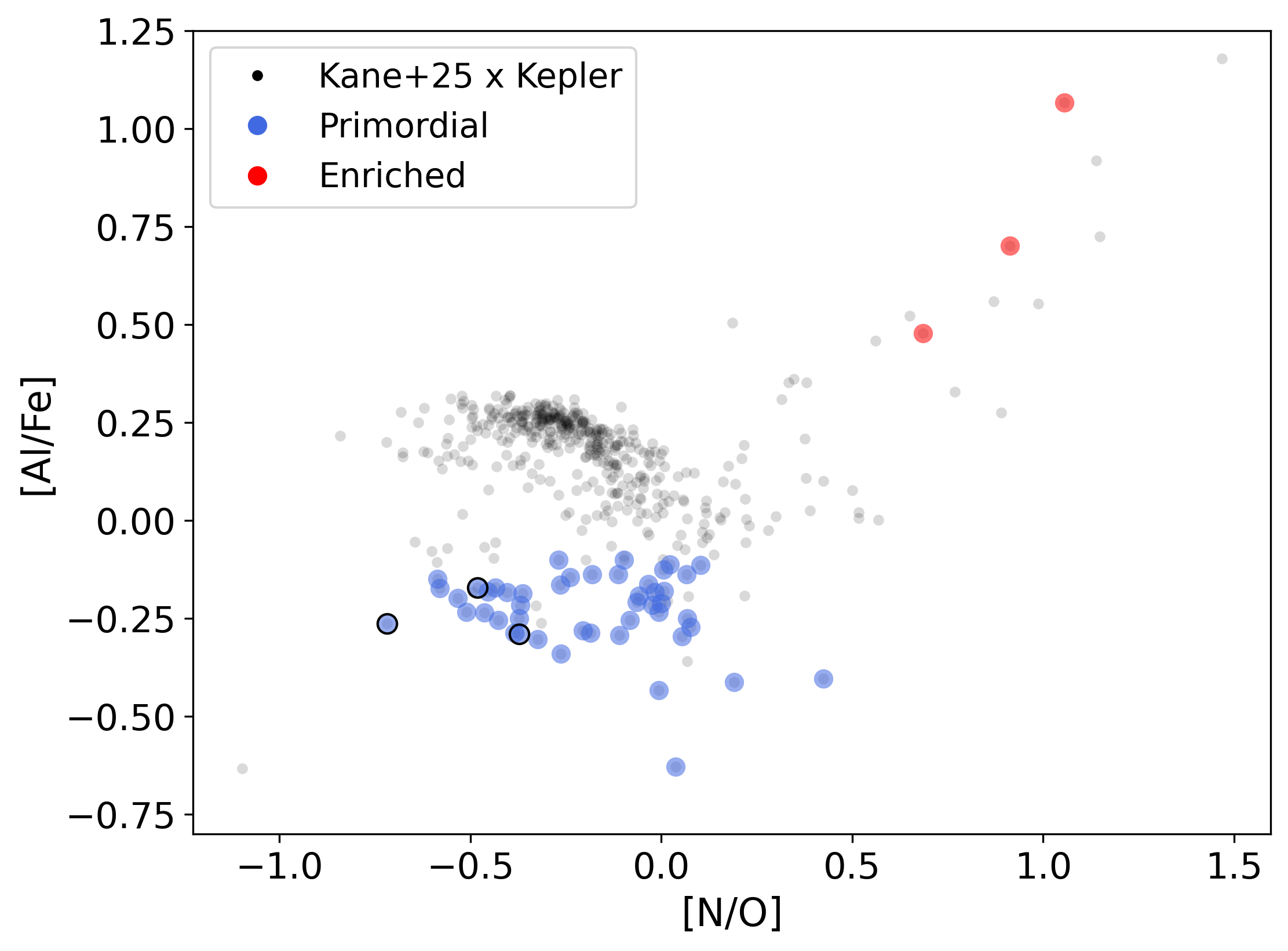}
    \caption{Primordial (blue) and enriched (red) stars identified in Kepler sample using abundances from the \textit{Gaia} BP/RP catalogue \citep{2025Kane} with [Fe/H] $<-0.5$ dex (black points). Stars in the Kepler sample with ages (three primordial stars) determined through asteroseismology are shown with black borders.}
    \label{fig:kepler_abundances_gaiabprp}
\end{figure}

In Section \ref{sec:data_gaia}, we briefly introduced the new abundance catalogue produced by \cite{2025Kane}, which added an extra 29 stars to our sample. The catalogue included predicted [N/O] and [Al/Fe] abundances, which we used to separate the stars into primordial and enriched. After cross-matching the new abundance catalogue with the \textit{Kepler} catalogue, we defined primordial stars as those with [N/O] $< 0.55$ and [Al/Fe] $< -0.1$, as shown as blue points in Figure \ref{fig:kepler_abundances_gaiabprp}. Enriched stars were then categorised by [N/O] $> 0.65$ and [Al/Fe] $> 0.1$ and are shown as red points in Figure \ref{fig:kepler_abundances_gaiabprp}. These criteria for primordial and enriched stars follow the general cuts outlined in \cite{2025Kane}, which were used for the same purpose of tagging primordial and enriched stars. The added sample of stars with new abundances were then cleaned to remove stars with unreliable astrometry as detailed in Section \ref{sec:data_gaia}, resulting in an additional three enriched stars and 26 primordial stars, three of which also have age estimates from asteroseismology as shown in Figure \ref{fig:kepler_abundances_gaiabprp}.

\section{Population classification using APOGEE DR17}
\label{sec:appendixPop}
In Section \ref{sec:data_pops}, we introduced the APOGEE abundances for which we classify stars in our sample as either primordial or enriched. In this appendix section, we will further describe and justify these classifications.

\begin{figure}[h]
    \centering
    \includegraphics[width=\columnwidth]{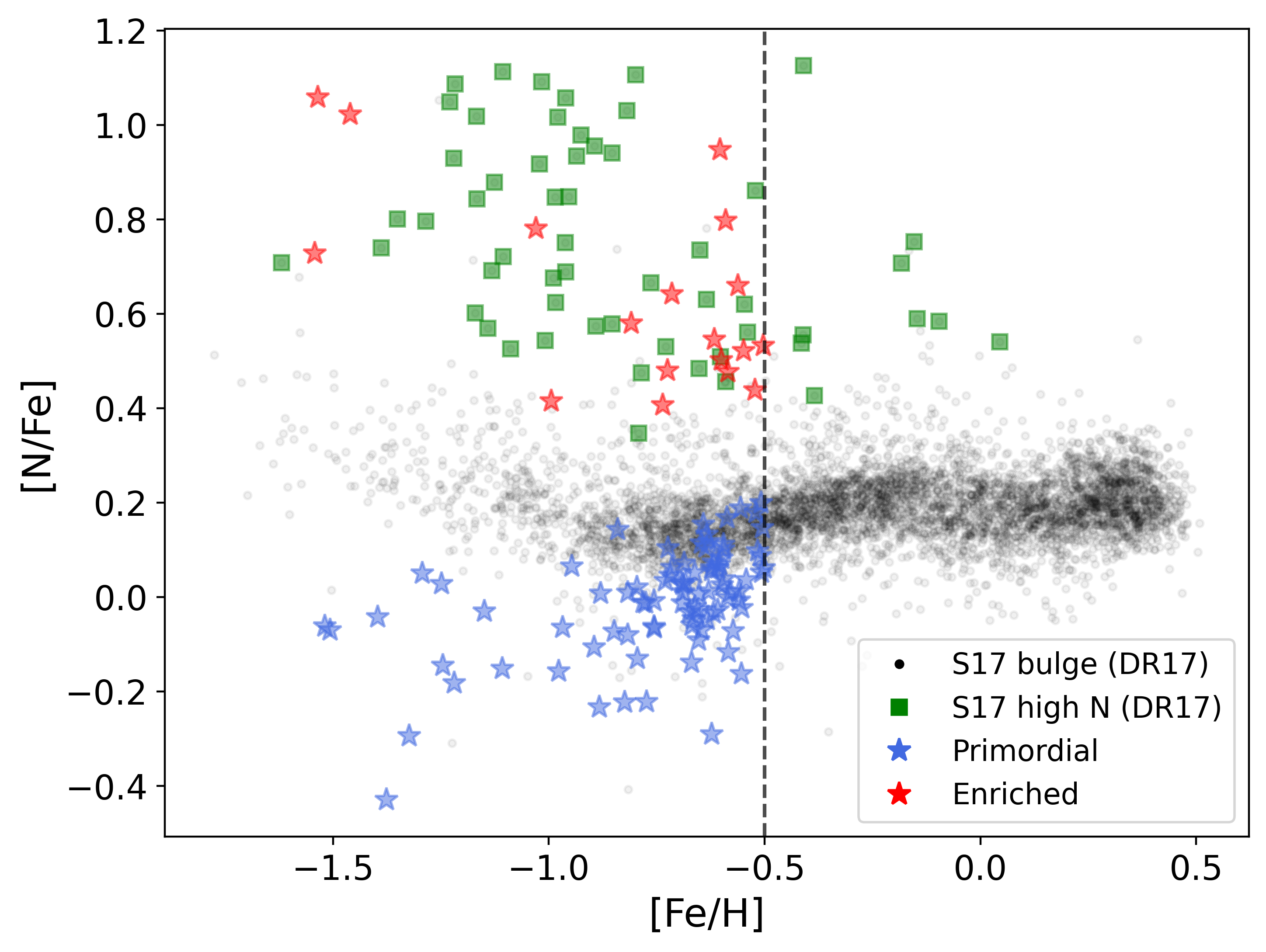}
    \caption{Nitrogen abundance versus metallicity for our primordial (blue stars) and enriched (red stars) samples, as well as the bulge stars (black points) and nitrogen-rich stars (green squares) presented in \cite{2017Schiavon} (S17 - updated from APOGEE DR12 to DR17 abundances). The vertical dashed line represents the metallicity limit of our sample ([Fe/H] $< -0.5$).}
    \label{fig:N_fe_S17}
\end{figure}

The abundance limits which define our enriched sample are consistent with the sample presented in \cite{2017Schiavon}, in which nitrogen-rich stars were selected as those with nitrogen abundances greater than $4 \sigma$ above the bulk of stars in the bulge, for the metallicity range $-1.8 \lesssim$ [Fe/H] $\lesssim 0.4$. This distinction between nitrogen-rich stars and bulge stars can most effectively be observed in Figure \ref{fig:N_fe_S17}, in which we include both the bulge (black) and nitrogen-rich (green) samples from \cite{2017Schiavon}, as well as our primordial (blue) and enriched (red) samples for comparison. We note that \cite{2017Schiavon} used APOGEE DR12 abundances, so in order to maintain consistency with our sample, we cross-matched both their bulge and nitrogen-rich sample with the newest APOGEE DR17 abundance catalogue, in order to remove the systematic differences between data releases. Despite the difference between our classification method and that of \cite{2017Schiavon}, our enriched sample shows consistency with their nitrogen-rich sample, mostly due to the metallicity regime of our sample. It becomes necessary to account for the increase in nitrogen exhibited by red giants for the metallicity regime [Fe/H] $> -0.5$, but as our sample lies in the [Fe/H] $< -0.5$ region, a simple cut in nitrogen is sufficient.

We note that APOGEE DR17 also provides information on the velocity dispersion (\texttt{VSCATTER}) of stars with multiple measurements of high SNR spectra, in which potential binaries are flagged as those with \texttt{VSCATTER} $>$ 1 km~s$^{-1}$. Within our sample, we have identified six stars which meet this criterium, including one enriched star (KIC 10796857) which is described further in \ref{sec:KIC10796857}, and five primordial stars. One of these primordial stars, KIC 10001167, has a \texttt{VSCATTER} $= 19.7$ km~s$^{-1}$ and is an old (10 Gyr), confirmed eclipsing binary hosting an oscillating RGB star \citep{2025AThomsen}. KIC 8914297 (\texttt{VSCATTER} $= 3.0$ km~s$^{-1}$) is a $G = $ 12.4 magnitude RGB star which has been flagged as a potential binary candidate (Mazzi et al., in prep), with a $G = 17.3$ magnitude neighbour star detected $< 3''$ away in Gaia DR3 \citep{2016gaia1,2022gaia2}. 
The remaining primordial stars are KIC 3553435 (\texttt{VSCATTER} $= 9.2$ km~s$^{-1}$), KIC 7265189 (\texttt{VSCATTER} $= 3.7$ km~s$^{-1}$) and KIC 10857579 (\texttt{VSCATTER} $= 1.1$ km~s$^{-1}$). Based on the low number of stars (6/133) which have \texttt{VSCATTER} $>1$ km~s$^{-1}$, we conclude that both the primordial and enriched samples are not biased towards stars which have obvious flags of binarity.

Finally, we note that the APOGEE DR17 catalogue provides both calibrated and uncalibrated versions of parameters such as $\rm{T_{eff}}$, log(g) and [Fe/H]. In order to derive the masses and hence ages of the enriched stars in our sample, the works of \cite{2021Montalban} and \cite{2026Willett} used the calibrated versions of these parameters, as the calibrated values minimise systematic biases. For the stars in our sample, and especially those listed in Table \ref{tab:kepler_sample}, the differences between calibrated and uncalibrated values for log(g) and [Fe/H] are negligible. However, there are discrepancies between calibrated and uncalibrated $\rm{T_{eff}}$ values of up to $\sim160$K for the stars in our enriched sample. We investigated the potential effects of these discrepancies for our sample, with the main concern being that this difference could increase the ages of the stars which are apparently too young to have originated from GCs. In the \cite{2021Montalban} method, individual mode frequencies dominate the fit for age determinations, rather than $\rm{T_{eff}}$. The ages determined through \texttt{AIMS} were therefore unburdened by the degree of difference between calibrated and uncalibrated $\rm{T_{eff}}$ values. Then, \cite{2021Miglio} found that a change of $100$K when using average seismic parameters to determine ages with \texttt{PARAM} results in a negligible change to the mean age and an increase of $\sim3\%$ in the intrinsic age spread. This is shown by comparing the results of the reference run R3 against R6 ($\rm{T_{eff}} + 100 K$) in Table 1 of \cite{2021Miglio}. As a worst-case scenario, we calculated the error propagation of a larger difference of $\rm{T_{eff}} = 160 K$ using the asteroseismic scaling relations, which uses inputs of $\Delta \nu$, $\rm{\nu_{max}}$ and $\rm{T_{eff}}$. We determined that this could affect the age estimation by up to $\sim15\%$ when using calibrated versus uncalibrated $\rm{T_{eff}}$ values.
However, we also note that the enriched stars in our sample with the largest discrepancies ($>100$K) are: KIC 5184073 (${12.2}^{+4.1}_{-3.1}$ Gyr), KIC 10398120 (${8.9}^{+5.4}_{-3.6}$ Gyr), KIC 8350894 (${6.4}^{+2.2}_{-1.8}$ Gyr) and KIC 3946701 (${1.4}^{+0.7}_{-0.4}$ Gyr), two of which already have ages consistent with the age distribution of MW GCs. For all enriched stars with ages that appear too young to have originated from GCs, a difference of $\leq 15\%$ in the asteroseismic age when using calibrated versus uncalibrated $\rm{T_{eff}}$ values is not sufficient to categorise them as old enough to have originated from GCs, therefore not affecting the main result of this work.

\section{Possible origins of individual enriched stars}
\label{sec:appendixB}
In this appendix we take a closer look at five of the enriched stars in our sample in order to investigate whether evidence suggests these stars originated from GCs, and if not, what the alternative origins could be. Currently, the alternative origin theory we will explore is that of enrichment through either past or current involvement in a mass-transfer binary system. Of the 13 enriched stars in our sample with asteroseismic ages, we explore the cases of KIC 5184073, KIC 10796857, KIC 8350894, KIC 8110538 and KIC 4920997 in more detail for the remainder of this section, using additional information from previous literature results where possible.

\begin{figure}[h]
    \centering
    \includegraphics[width=\columnwidth]{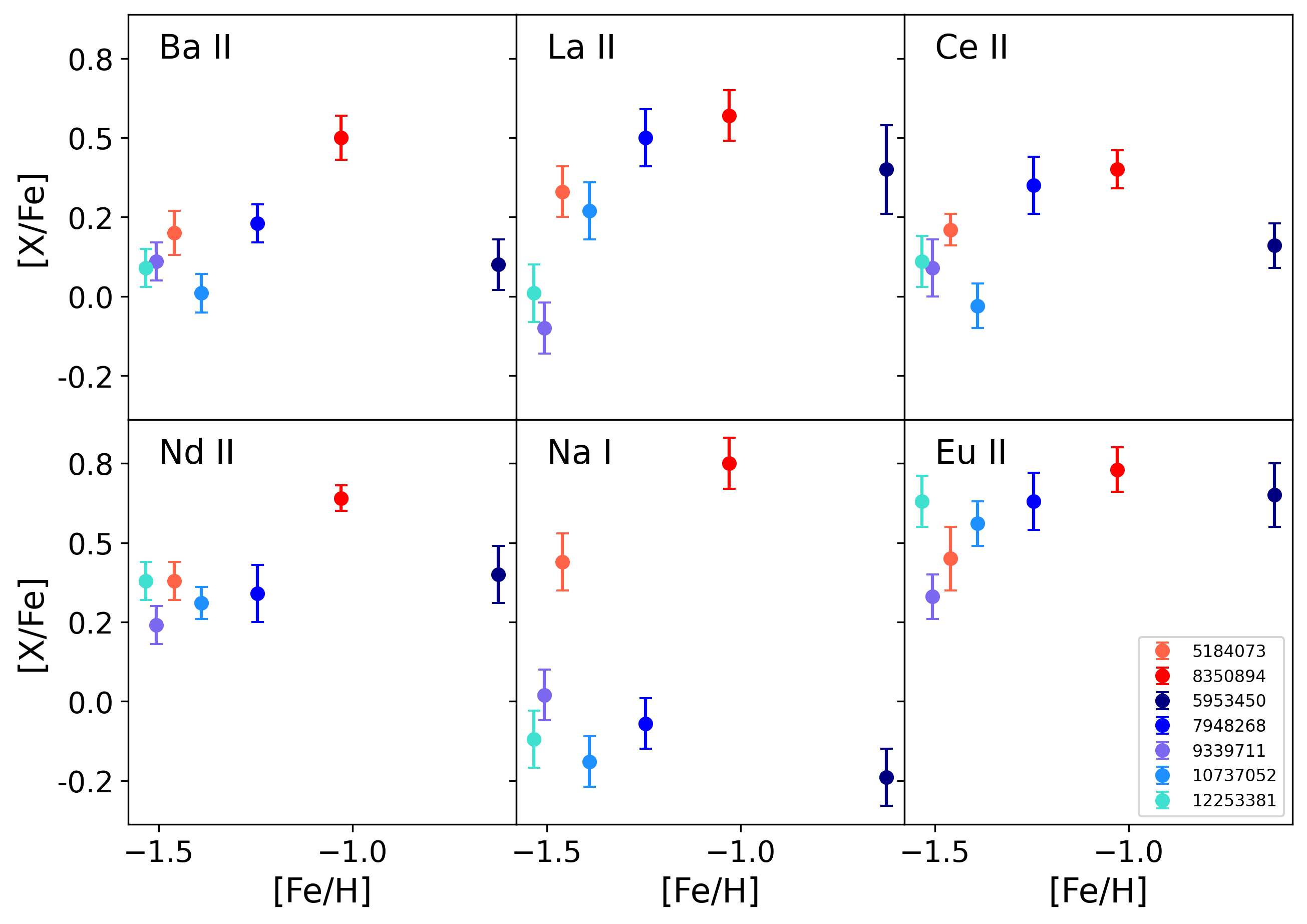}
    \caption{Abundances of two enriched (shades of red) and five primordial (shades of blue) stars in our sample with abundances obtained by \cite{2021Matsuno}.}
    \label{fig:matsuno_abundances}
\end{figure}

We will refer to the abundances of halo red giants which were derived from high resolution optical spectra from the High Dispersion Spectrograph on the Suburu Telescope by \cite{2021Matsuno}. Our sample has seven RGB stars in common with \cite{2021Matsuno}. This cross-match added many elements which can be used as indicators of binary mass transfer. For example, a red giant star with an AGB companion can show enrichments in elements such as barium (Ba), lanthanum (La), cerium (Ce) and neodymium (Nd), while a neutron star/black hole companion could be indicated by enrichments in europium (Eu), which are not available in the APOGEE DR17 abundances. However, while enrichments in these specific elements is potentially a sign of binary mass transfer between RGB and AGB stars, this is only one of many combinations in terms of binary pairs. Mass transfer in a binary system will not necessarily leave traceable evidence in terms of specific chemical enrichments, so we are only able to chemically trace those which do leave evidence. A lack of these specific enrichments in any star in our sample therefore does not necessarily rule out binarity. In Figure \ref{fig:matsuno_abundances}, we show the abundances from \cite{2021Matsuno} for the two enriched stars and five primordial stars, in terms of the elements mentioned above, as well as the sodium (Na) abundance, confirming that these two enriched stars which were selected due to enrichment in N and Al from APOGEE DR17, are also enriched in Na, in comparison to the primordial stars which are all Na-poor.

\subsection{KIC 5184073}

KIC 5184073 has an asteroseismic age of $12.2^{+4.1}_{-3.1}$ Gyr, which is mostly consistent with the expected age distribution of MW GCs (see the upper right panel of Figure \ref{fig:kepler_sample}), but with substantial enough errors on the lower age estimate to also push the star into the `too young to have originated from a GC' age range. This enriched RGB star has orbital properties consistent with the thick disc, while its orbital energy of $E = 1.6 \times 10^5$ [$\rm{km^2/s^2}$]) and angular momentum $L_z = 734$ [km/s kpc] could be indicative of an accreted origin. It also lacks enrichment in terms of Ba, La, Ce and Nd \citep{2021Matsuno}. Enrichment in these elements is strong evidence of mass transfer with an AGB companion, but we also note that a lack of enrichment in these elements does not necessarily rule out mass transfer. We conclude that for this star, the evidence suggests this may be a star which originated and was enriched within a GC before being ejected or dissolved to the thick disc.

\subsection{KIC 10796857}
\label{sec:KIC10796857}
The asteroseismic age of this enriched RGB star is $1.9^{+0.1}_{-0.2}$ Gyr, which is not consistent with the age of any MW GC, indicating this is extremely unlikely to be its origin. The asteroseismic age of this star has been determined using both methods outlined in Section \ref{sec:data_seismo}, producing consistently young ages. For example, we quote the age estimate using \texttt{AIMS} \citep{2021Montalban}, but the star was also analysed using \texttt{PARAM} with $\Delta \nu$ as an input, producing an age estimate of $1.6^{+0.4}_{-0.1}$ Gyr, as well as using luminosity as an input, producing an age of $2.2^{+0.3}_{-0.3}$ Gyr, showing consistency across all age estimation methods. This star is located in the halo, with its orbital energy of $E = 1.8 [\rm{km^2/s^2}]$ combined with a positive value of $L_z = 205$ [km/s kpc], indicating prograde motion, and therefore an in-situ origin.

The release of APOGEE DR17 included the velocity dispersion of targets that were observed multiple times: \texttt{VSCATTER}, in which a target star with \texttt{VSCATTER} $>$ 1 km~s$^{-1}$ from multiple visits of high SNR visit spectra can indicate that the target is a binary. For KIC 10796857, the \texttt{VSCATTER} $=1.5$ km~s$^{-1}$, with RV $ = -222.73, -220.53, -219.56$ (all with uncertainties $\pm 0.04$) km~s$^{-1}$, from three visits in May of 2013, showing indications of a binary system.

Additionally, with the newest release of \textit{Gaia} DR3, an additional parameter: \texttt{non\_single\_star} was introduced, in which a non-zero value indicates either an astrometric, spectroscopic (single-lined SB1; and double-lined SB2) or eclipsing binary \citep{2023Gaia}. KIC 10796857 has a \texttt{non\_single\_star} value of $2$, indicating that it has been flagged as a spectroscopic binary SB1 with a period of $149.1 \pm 0.5$ days, based on radial velocity variations. This evidence of binarity from both APOGEE DR17 and \textit{Gaia} DR3, together with an asteroseismic age of $1.9^{+0.1}_{-0.2}$ Gyr, support the conclusion that this star was enriched through mass transfer with a companion, in such a way as to mimic the same enrichment patterns observed in GC enriched stars. The combination of confirmed binarity and an asteroseismic age of $1.9^{+0.1}_{-0.2}$ Gyr supports the idea that the star has undergone binary mass-transfer.

\subsection{KIC 8350894}
KIC 8350894 is an RGB star located in the Galactic halo, but with higher orbital energy ($E = 1.6 \times 10^5$ [$\rm{km^2/s^2}$]) and lower angular momentum ($L_z = 436$ [km/s kpc]), in comparison to stars formed in-situ in the MW, suggesting it could have had an accreted origin. This star is also in common with \cite{2021Matsuno}, allowing insight into additional chemical abundances such as Eu, Na, Ba, La, Ce and Nd. In Figure \ref{fig:matsuno_abundances}, we show that KIC 8350894 is also enriched in Na - a distinctive chemical enrichment which, when paired with the depletion in O, validates the selection of this star as similar in chemistry to an enriched GC star. However, this star is also enriched in every element expected for binary mass transfer with an AGB star, that is: Ba, La, Ce and Nd at levels $>0.5$ dex, as seen in Figure \ref{fig:matsuno_abundances}.

This star was also included in the sample of \cite{2024Xu}, in which they use full action distributions in combination with metallicities and abundances from LAMOST to trace the origins of halo field RGB stars. They analysed KIC 8350894, giving a confidence value of $p = 0.36$ that it once belonged to the 12.7 Gyr old GC NGC 6352; however, due to the low confidence value, they did not conclude NGC 6352 as the origin. The apparent asteroseismic age of this star is $6.4^{+2.2}_{-1.8}$ Gyr, but as this star shows chemical signatures of mass-transfer with an AGB companion, this age estimation could be affected by the assumption of single stellar evolution models. We conclude that this star is most likely the product of mass-transfer with an AGB companion, but as the asteroseismic age is affected by this process, we cannot rule out that the star originated in a cluster, as it is also possible that it underwent mass-transfer with an AGB companion while within a cluster, before being ejected into the field.

\subsection{KIC 8110538}
We present here the case of a CHeB thick disc star with kinematics consistent with in-situ formation.
The asteroseismic age determined for this star using \texttt{PARAM} with the $\Delta \nu$ constraint (see Sec.~\ref{sec:data_seismo}) is $5.6^{+1.9}_{-1.3}$ Gyr ($M=1.10^{+0.10}_{-0.10} \rm{M_\odot}$); however, the age estimate changes to $1.9^{+0.5}_{-0.4}$ Gyr ($M=1.59^{+0.15}_{-0.12} \rm{M_\odot}$) using the luminosity. The difference in the masses derived with the two sets of parameters indicates that the constraints are not compatible, and in particular the luminosity of the star could suffer from contamination by a nearby star that is not resolved with 2MASS photometry \citep{2025Mazzi}. These aspects will be explored further in a follow-up investigation.

\subsection{KIC 4920997}
\label{sec:KIC4920997}
This enriched CHeB star is located in the thick disc and exhibits kinematics consistent with in-situ formation within the MW. The asteroseismic age of $18.1^{+1.1}_{-1.9}$ Gyr results from the mass estimate of $0.77^{+0.02}_{-0.01} \rm{M_{\odot}}$ using \texttt{PARAM} with the $\Delta \nu$ constraint. Since this is a low-mass CHeB star that went through a helium flash, the derived age is highly dependent on the RGB mass loss assumed in the models. The \texttt{PARAM} grid used in the present case assumed $\eta=0.2$, which is less than inferred from observations at this metallicity \citep[e.g.][]{2024Brogaard}. The adoption of a mass loss in line with other stars at the same mass and metallicity would likely result in an age, which is still old, but not older than the universe. Thus, the star could be consistent with having a GC origin. However, the star could also be younger, but having lost a larger-than-average amount of mass during evolution in a mass-transfer binary similar to the evolution of KIC\,4937011 suggested by \cite{2024Matteuzzi,2017Handberg}.

\newpage
\onecolumn
\section{Table data}
\label{sec:appendixC}

\begin{table*}[h]
\centering
\label{tab:kepler_sample}
\caption{Asteroseismic properties of the enriched stars in our sample, with adopted metallicity and effective temperature inputs, as well as nitrogen abundances from APOGEE DR17.\protect\footnotemark}

\resizebox{\hsize}{!}{
\begin{tabular}{cccccccc}
\hline
KIC      & Evolutionary state & Age [Gyr]              & Mass [$M\odot$]           & [Fe/H]           & $\rm{T_{eff}}$ [K] & [N/Fe]          & APOGEE ID          \\
\hline
1726211  & CHeB                 & ${3.4}^{+0.9}_{-1.0}$  & ${1.28}^{+0.17}_{-0.10}$  & -0.60 $\pm$ 0.05 & 4973 $\pm$ 50 & 0.50 $\pm$ 0.02 & 2M19300107+3717340 \\
2165615  & CHeB                 & ${1.8}^{+0.5}_{-0.3}$  & ${1.56}^{+0.11}_{-0.13}$  & -0.74 $\pm$ 0.05 & 5000 $\pm$ 50 & 0.41 $\pm$ 0.03 & 2M19302512+3730446 \\
3120676  & CHeB                 & ${1.2}^{+0.3}_{-0.4}$  & ${1.90}^{+0.38}_{-0.19}$  & -0.55 $\pm$ 0.05 & 5053 $\pm$ 50 & 0.52 $\pm$ 0.02 & 2M19293022+3816119 \\
3850744  & CHeB                 & ${0.8}^{+0.2}_{-0.1}$  & ${2.22}^{+0.17}_{-0.20}$  & -0.59 $\pm$ 0.05 & 5157 $\pm$ 50 & 0.48 $\pm$ 0.03 & 2M19261974+3859443 \\
3946701  & RGB                & ${1.4}^{+0.7}_{-0.4}$  & ${1.69}^{+0.22}_{-0.19}$  & -0.56 $\pm$ 0.05 & 4838 $\pm$ 50 & 0.66 $\pm$ 0.03 & 2M19170711+3904391 \\
4920997  & CHeB                 & ${18.1}^{+1.1}_{-1.9}$ & ${0.77}^{+0.02}_{-0.01}$  & -0.60 $\pm$ 0.05 & 4920 $\pm$ 50 & 0.95 $\pm$ 0.03 & 2M19233926+4003386 \\
5184073  & RGB                & ${12.2}^{+4.1}_{-3.1}$ & ${0.79}^{+0.08}_{-0.06}$  & -1.46 $\pm$ 0.05 & 4880 $\pm$ 50 & 1.02 $\pm$ 0.05 & 2M19233833+4018284 \\
8110538  & CHeB                 & ${5.6}^{+1.9}_{-1.3}$  & ${1.10}^{+0.10}_{-0.10}$  & -0.59 $\pm$ 0.05 & 4975 $\pm$ 50 & 0.80 $\pm$ 0.03 & 2M19442885+4354544 \\
8350894  & RGB                & ${6.4}^{+2.2}_{-1.8}$  & ${0.98}^{+0.10}_{-0.08}$  & -1.03 $\pm$ 0.05 & 4907 $\pm$ 50 & 0.78 $\pm$ 0.03 & 2M19004420+4421082 \\
8620063  & CHeB                 & ${1.1}^{+0.1}_{-0.3}$  & ${1.89}^{+0.23}_{-0.11}$  & -0.52 $\pm$ 0.05 & 5149 $\pm$ 50 & 0.44 $\pm$ 0.03 & 2M19202332+4442142 \\
10398120 & RGB                & ${8.9}^{+5.4}_{-3.6}$  & ${0.87}^{+0.16}_{-0.11}$  & -0.99 $\pm$ 0.05 & 4775 $\pm$ 50 & 0.41 $\pm$ 0.03 & 2M19160273+4735390 \\
10796857 & RGB                & ${1.9}^{+-0.1}_{-0.4}$ & ${1.51}^{+0.12}_{--0.03}$ & -0.73 $\pm$ 0.05 & 5040 $\pm$ 90 & 0.48 $\pm$ 0.03 & 2M19264357+4807193 \\
11443139 & CHeB                 & ${1.9}^{+1.1}_{-0.8}$  & ${1.55}^{+0.36}_{-0.23}$  & -0.72 $\pm$ 0.05 & 5023 $\pm$ 50 & 0.64 $\pm$ 0.03 & 2M18584348+4923267 \\
\hline
\end{tabular}
}
\end{table*}

\footnotetext{Asteroseismic masses and ages for these 13 enriched stars were determined using \texttt{PARAM}, except for KIC 10796857 using \texttt{AIMS}.\\
Age $> 13.0$ Gyr for KIC 4920997 likely due to underestimated RGB mass-loss efficiency at low metallicity (see Section \ref{sec:KIC4920997}).}

\end{appendix}

\end{document}